# Analysis of Dead Reckoning Accuracy in Swarm Robotics System

Weihang Tan and Timothy Anglea and Yongqiang Wang

*Abstract* —The objective of this paper is to determine the position of a single mobile robot in a swarm using dead reckoning techniques. We investigate the accuracy of navigation by using this process. The paper begins with the research background and social importance. Then, the specific experimental setup and analysis of experimental results are presented. Finally, the results are detailed and some potential improvements are provided.

## I. Introduction

ACCURACY, speed, and safety are the most important factors of the robots. At present, controlling the correct heading and position is a very critical problem for the automation of cars and in some other mobile robots. In some other semi-automatic systems, robots or cars can perform well since they have an artificial modification of their heading or velocity. But in the swarm robotic, semi-automation is not achievable because of a heavy load of robots. So, we need them to do the self-modification and self-navigation when they are moving. This research booms since artificial intelligence allows for great progress in the automation-control and intelligent system area. For example, companies like Google and Tesla Motors offer a lot of funding on investigating the autonomous cars and autonomous mobile robots, and so the dead reckoning algorithm is applied into the navigation of autonomous automobiles and robots.

Dead reckoning allows a navigator to determine its present position by projecting its past courses steered and speeds over ground from a known past position [1]. We also can use this technology in our swarm robots to determine their position by themselves without the need of additional sensors, using only encoders that sense the distance traveled by each wheel. By calculating the distance traveled and relative heading changes each time data is received, an estimate of position can be formed. To calculate a new estimate, the navigator updates is current estimated position by the distance calculated, ß, in the direction previously determined, μ, and changes its relative heading to be used in the next iteration. As shown in Fig. 1, the estimated position tracks the actual position. Better accuracy is achieved as the rate of data reception is increased.

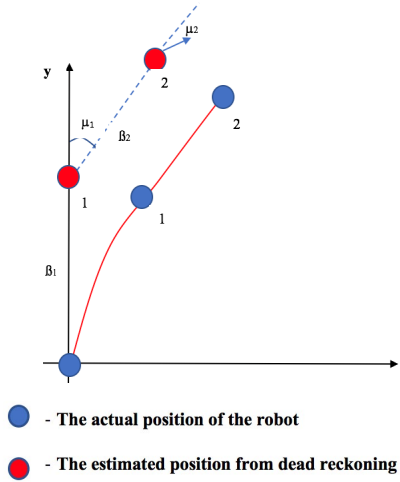

**Fig. 1: Model of Dead Reckoning Corresponding Times of Data Reception.**

The Riemann integral can be used to partition a robot's trajectory in a very small piece of distance. We can use angle modification to correct robots' direction in each small piece of distance by using dead reckoning. Localization is very important to know the current location, which is the blue circle in Fig. 1, and robots can know the current location by using the dead reckoning algorithm. Particularly, the electric compass can be used to determine the heading by our robots.

We use time instead of distance as our parameter to partition into a small piece interval. In each time interval, robots can calculate the current location and heading, then do the angle modification if it is necessary. The trajectory in the small-time interval might not be linear or have an incorrect heading, but, if it uses the Riemann integral to sum up all the distances in these specific times, it can theoretically perform a linear trajectory and overall correct heading since it can modify its direction in each small movement.

In section II, we discuss the experimental setup of our experiment and some computations which are used to convert the reading data from the robot to particular data used for our experimental accuracy analysis. Then, in section III, we talk about results of the experiments and some challenges that happened in the experiment, and offer our conclusions in section IV. Based on the challenges in section III, section V will introduce some ideas for future work which may help to reduce some negative effects and help to improve the performance of the robots.

Weihang Tan, Timothy Anglea and Yongqiang Wang are with Department of Electrical and Computer Engineering, Clemson University, Clemson, SC 29634 USA (e-mail: {wtan, tbangle, yongqiw}@clemson.edu).

## II. Experimental Setup

The swarm robotic system produces the position and heading by using the closed-loop control system. We use the iRobot Create 2 Roomba 600 (Roomba) as our experiment target and robotic platform. Also, we investigate the heading, position and velocity sub-system of the project.

### A. User Interface:

We used our computer terminal as our user interface and gave some specific commands for testing. Also, we used control variation method to investigate the dead reckoning algorithm's performance with speed and time in Table I.

TABLE I. TESTING VARIAVLES OF ROOMBA

|        | Speed [mm/s] | Time [seconds] |
|--------|--------------|----------------|
| Test 1 | 75           | 65             |
| Test 2 | 75           | 65             |
| Test 3 | 100          | 41             |
| Test 4 | 100          | 52             |
| Test 5 | 50           | 91             |
| Test 6 | 50           | 90             |
| Test 7 | 150          | 39             |

### B. Electric Compass:

An electric compass can help our robots determine the heading, which is the feedback branch of our close-loop control system. In this experiment, we use a HMC5883L Triple Axis Magnetometer Breakout Board to achieve our specific purpose. We can calculate the heading angle by using the x and y outputs from the magnetometer. Once we get the heading of the robots, the robots can do the navigation by themselves. Generally, it only helps robots' navigation purpose, but not in the dead reckoning algorithm.

### C. I/O Interface:

The Arduino UNO microcontroller board was used as our I/O interface between users and the actuator platform, which is the Roomba robot.
The Arduino has a small volume, and it can be powered by the Roomba for operation. Also, the electric compass can use the $I^2C$ communication bus of the Arduino. On the other hand, the Arduino did the dead reckoning algorithm by programming and angle modification as well as position computation.

### D. Actuators:

The Roomba was our actuator platform (specific robotic platform). Since we need to investigate the heading and position, wheels and geometry of the Roomba are decisive factors. For our programming required, the Roomba encoder transfers the data to the Arduino and it can print data as human-readable ASCII text about the heading degree from the electric compass and counts of left & right wheels from the Roomba. In the dead reckoning algorithm, measured distance (ß) (unit: mm) and modified angle (µ) (unit: degree) in Fig. 1 need to use counts of the left & right wheel to convert to mm and degree for analysis. By achieving this purpose, the following mathematical model need to be applied.

From the side view of the Roomba in Fig. 3, we know the diameter of the wheel is 72mm, so the Roomba travels $72\pi$ mm per wheel revolution. Then, given that the encoder gives 508.8 counts/revolution, we get approximate conversion between counts and distance. The length converted constant is denoted as $d$ and equation as following:

$$d = 508.8 \div 72\pi \approx 2.25 \; [counts/mm] \qquad (1)$$

From the bottom view in Fig. 3, the distance between the left and the right wheel is 235mm and there is 360° per revolution (each wheel travels 360° per revolution of the Roomba). Thus, the circumference that the Roomba rotates 360° in the same place is $470\pi$. Then, we need to use equation (1) to convert counts to radian, giving us the angular constant, denoted Ø.

$$\emptyset = 2 \times 235\pi \times 508.8 \div 72\pi \approx 9.226 \; [counts/rad] \qquad (2)$$

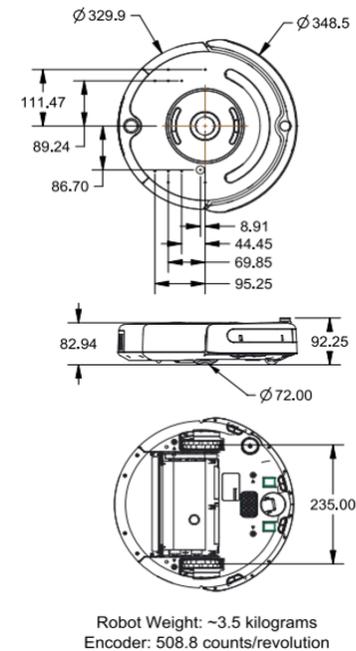

Fig. 2. Schematics of Roomba [2]

### E. Experimental Computation:

Once we received the data from the encoder, we could calculate the actual position, measured distance, as well as speed, and analyze the heading. We can get the printed data of the serial port on the Arduino, and record data in each step (Δt). Since we need to calculate the actual position, we set the plane as a rectangular coordinate system and, for easier calculation, the initial position as (0,0).

- Counts:

Given the $n^{th}$ left and right wheel encoder counts, the difference between the current and previous counts, ($\Delta RW_n$ for right wheel, $\Delta LW_n$ for left wheel), can be found as follows:

$$\Delta RW_n = RW_n - RW_{n-1} \quad (3)$$

$$\Delta LW_n = LW_n - LW_{n-1} \quad (4)$$

- Measured Distance:

To calculate the measured distance (ß) that the Roomba runs, we need the average of the left and right wheel differences:

$$\beta_n = [(\Delta RW_n + \Delta LW_n) \div 2] / d \quad [unit: mm] \quad (5)$$

- Modified Angle:

Moreover, we need to calculate the modified angle (μ) which is associated with the heading of our algorithm:

$$\mu_n = [(\Delta RW_n - \Delta LW_n) \div \varnothing] \cdot \pi \div 180° \quad [unit: °] \quad (6)$$

- Position:

Since we set Roomba run in a rectangular coordinate system, the $n^{th}$ x-axis value ($x_n$) and y-axis value ($y_n$) are calculated from the modified angle μ (in degree) as well as the measured distance ß (in mm):

$$x_n = x_{n-1} + \beta_n \times sin(\mu_n) \quad (7)$$

$$y_n = y_{n-1} + \beta_n \times cos(\mu_n) \quad (8)$$

- Speed:

Speed (*v*) can be calculated in terms of measured distance and the time step (Δt).

$$v = \frac{\beta}{\Delta t} \quad [unit: mm/s] \quad (9)$$

### III. EXPERIMENTAL RESULTS

Experiments were performed in an outdoor environment since an indoor environment has electromagnetic interference (EMI), which may influence the performance and function of our electric compass.

In each test, we used three distances (result shown in Table II) to determine whether the robots follow the commanded speed or not, and named the three distances as actual distance, theoretical distance and measured distance, respectively. We measured the actual distance from the original position and the final position where the robots finally stop. The theoretical distance is the distance based on the speed we commanded the robot in Table I. The measured distance is from the calculation of the Roomba's wheel encoders. After analysis, we can see that the measured distance and the actual distance are almost the same, but the theoretical distance is significantly different from the actual or measured distance.

TABLE II. ACTUAL, THEORETICAL AND MEASURED DISTANCE OF ROOMBA

|  | Actual Distance [m] | Theoretical Distance [m] | Measured Distance [m] |
|---|---|---|---|
| Test 1 | 5.83 | 4.875 | 5.804 |
| Test 2 | 5.79 | 4.875 | 5.703 |
| Test 3 | 4.54 | 4.1 | 4.516 |
| Test 4 | 5.73 | 5.2 | 5.934 |
| Test 5 | 5.80 | 4.55 | 5.797 |
| Test 6 | 5.65 | 4.5 | 5.815 |
| Test 7 | 5.70 | 5.85 | 5.641 |

We also used the visual inspection method to measure the rotation offset of the Roomba. Then, we found that it had drifted from the initial heading position in all tests. Specifically, it had a north-east drift in Test 1.

With the limited amount of time for this project in this semester, we only use one robot, and tested in the outdoor environment, but not in an indoor environment. Also, we have some hardware and software challenges to be solved in future semesters.

#### A. Linear:

We apply the serial printed data from the Arduino, equations in experimental computation and use MS Excel for data analysis in Test 1. The original x, y-coordinate diagram is shown in Fig. 3. For linear trajectory analysis, linear regression of the system needed to be considered, as depicted in Fig. 4, and we used Excel to calculate explained variation (total variation) $R^2$.

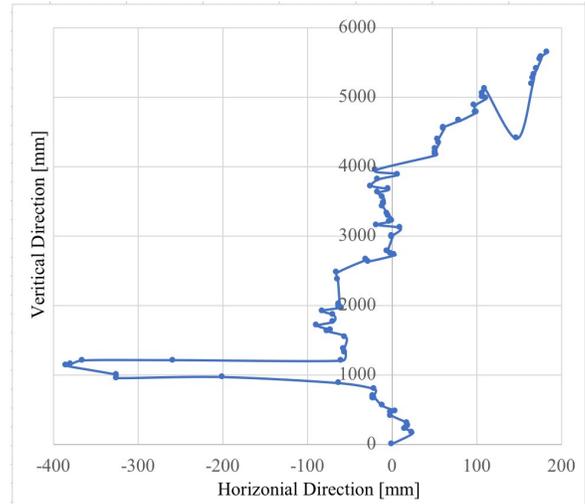

Fig. 3. Linear Analysis in x, y Coordinate System of Test 1. This shows the partial linear performance and measured trajectory of our robot.

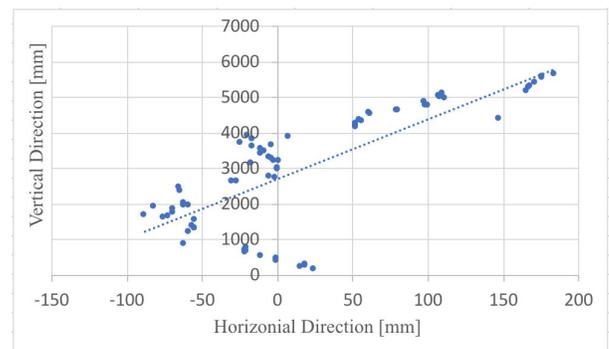

Fig. 4. Linear Regression Analysis of Test 1. It shows that those acceptable data points are regress in strict line on first quartile.

In the diagram in Fig. 3, we can see a group of the discrete positions trend to the north-east direction in the first quartile of the diagram, which is the same as we observed. With the analysis of the group of points, we can find that some reading errors occurred that made the heading be different. In other words, the erroneous data of the left and right wheel encoder also will make an error in our modified angle μ$_n$ and the future position will also be influenced by the reading error. In the linear regression analysis (shown in Fig. 4), after we remove all the unacceptable data, the $R^2$ is only 0.5783, which shows a not very linear performance of the track. From the linear fit function (y = 16.828x + 2705), we can predict that the robot will keep running to the north-east in the future sample time. Based on these two diagrams and analysis, we can consider our system to be partially linear.

### B. Error Propagation (Heading Error):

Error propagation is a common weakness of the dead reckoning algorithm. The error might come from an encoder reading error, ground influence, etc. In the experimental computation part, we know that the current x and y values are based on the previous left & right wheel encoder value, so once a reading error occurs, it will propagate to the future values. After we read our data table, we can find the error propagation in our system, which is depicted in Table III. From the table high-lighted part, we can see the x-axis values jump from -61.915 to -200.543 in two adjacent data points and the next five values are propagating the previous error value in number 103 in the table. Based on this reason, the heading of our robot cannot run in a straight direction strictly and make the inaccuracy of speed. This is also one of the reasons why the Roomba drifted to the north-east (right) trend on Test 1, and the theoretical & measured distance do not match.

TABLE III.  ILLUSTRATION OF PROPAGATED ERROR DATA

| Data Number | Right Wheel Encoder Counts | Left Wheel Encoder Counts |
| --- | --- | --- |
| **102** | **-61.915** | 873.609 |
| **103** | **-200.543** | 962.477 |
| 104 | -325.060 | 948.201 |
| 105 | -324.717 | 993.533 |
| 106 | -384.548 | 1123.538 |
| 107 | -379.684 | 1146.358 |
| 108 | -365.296 | 1146.358 |

### C. Noise of the Environment:

Noise is a very critical influential factor in our system. The electric compass is influenced by EMI. Before the tests, we have tried to see how much EMI will influence our electric compass sensor, so we opened the compass app on a cellphone, and we put the cellphone close to the magnetometer. Then we found that the reading of the magnetic x, y, and z coordinates from the serial printed data became chaotic, so that we can consider the compass or devices with a compass may have interference to the electric compass sensor. Considering the indoor environment is a laboratory building, which is full of electrical devices, and our lab environment cannot perform as a Faraday cage which can reduce the influence of EMI, we do the tests in an outdoor environment.

### D. Velocity Fluctuation:

Velocity fluctuation influenced the accuracy of the Roomba. Since we tested the Roomba on uneven and rough outdoor ground, which is different from the floor tile in a room, the velocity may be influenced by the ground conditions. If the velocity changes, the acceleration '*a*' also changes because of the equation of movement illustrated as following:

$$\vec{a} = \left(\vec{v} - \vec{v_o}\right) / \Delta t \qquad (10)$$

Once the previous velocity changes, the acceleration will also be changed. Moreover, the classical mechanics tells the relationship between force **F** and acceleration *a*, so that the force on the Roomba made its actual heading angle different from the desired one.

Based on this problem, we consider that the actual distance does not match the theoretical distance because of the influence of velocity fluctuation. The Roomba cannot move uniformly and have different velocities in each step, so that the entire actual speed is different from the commanded speed and this makes the actual distance different.

## IV. CONCLUSION

In our experiment, using the dead reckoning algorithm still has the heading and distance problem, but overall it is accurate enough to apply in the robots. After we solve these problems, we can apply the dead reckoning algorithm in a wider range in robotics.

In future semesters, we will improve our robots based on some shortcomings in the experimental results. Also, we will test more robot samples, and especially redo our test in several robots working synchronously. On the other hand, some specific software and hardware development are introduced as follows.

## V. FUTURE WORK

### A. Error Scanning:

For solving the error propagation problem, we can set a threshold of values for the encoders of both the left and right wheels. If the Arduino receives some values which exceed the threshold value, the Arduino will require the Roomba to send new values until they are under the acceptable threshold value. Since we analyzed the encoder data in the left and right wheels, we can know that the encoder reading cannot increase 300 counts in each 0.5 second step. We could set the threshold value as positive 300 plus the previous reading value for each 0.5 second. This way is easier to implement in code and does not require additional hardware or extensive computation, so it will be a priority.

### B. Inertial Navigation System:

Generally, an inertial navigation system consists of a gyroscope sensor and an accelerometer sensor. These two sensors can improve the heading and velocity measurements.

A gyroscope sensor has a very good fixed-axis movement performance in the mobile robot. Since the electric compass can detect the specific magnetic heading for our robots, the gyroscope can use the heading angle as the axis, then it can fix our robots running in this axis, which can solve our heading problem. Also, the error propagation of the dead reckoning algorithm is reduced by using this gyroscope sensor.

An accelerometer sensor can keep the direction of trajectory. Changes in lateral acceleration indicate a change in direction due to uneven ground conditions. These changes can be detected and used to correct the trajectory by using the accelerometer sensor. After we apply this sensor, we can solve the velocity fluctuation problem, and the actual distance will match the theoretical distance.

*C. Control System:*

A Proportional–Integral–Derivative Controller (PID controller) and Lead-Lag compensation can reduce the error propagation problem. The compensator can do some calculations to reduce the steady-state error so that the propagation problem can be improved. The compensator improvement can be done in Simulink software.

In PID Control, since the error signal is inversely proportional to the proportional gain, K, we can just increase the K value so the error can be reduced. Also, the integral controller can guarantee the error will be zero, which can make the system more stable.

In a Lead-Lag Compensator, the lag compensation can add a pole and a zero to reduce the error in the system and make the system performance efficiently.

*D. Algorithm Filter:*

The Kalman filter has numerous applications in technology. A common application is for guidance, navigation, and control of vehicles, particularly aircraft and spacecraft [3]. We can use a Kalman Filter to reduce the noise and inaccuracies of our system and the heading problem, and we can use MATLAB to model our dynamic system and apply the algorithm.